\documentclass[preprint,10pt]{elsarticle}

\journal{Physica A}

\usepackage{graphicx,bbm,setspace,amssymb,amsfonts,amsmath,mathtools,mathrsfs,stmaryrd,amsthm,bm,etoolbox,comment,hyperref,url,caption,subcaption,color,enumitem,algorithm,booktabs,microtype,nicefrac,lingmacros,nccmath,tree-dvips,tikz-cd}
\usepackage[noend]{algpseudocode}
\usepackage{algorithm}

\usepackage[colorinlistoftodos]{todonotes}
\usepackage[utf8]{inputenc} 
\usepackage[T1]{fontenc}    

\usepackage[noend]{algpseudocode}

\newtheorem{thm-defn}[theorem]{Theorem/Definition}

\theoremstyle{definition}

\theoremstyle{remark}

\DeclareMathOperator{\E}{\mathbb{E}}

\newcommand{\rvline}{\hspace*{-\arraycolsep}\vline\hspace*{-\arraycolsep}}

\newcommand*{\vertbar}{\rule[-1ex]{0.5pt}{2.5ex}}
\newcommand*{\horzbar}{\rule[.5ex]{2.5ex}{0.5pt}}
\newcommand{\ignore}[1]{}{}

\begin{document}

\begin{frontmatter}

\title{On the systemic nature of global inflation, its association with equity markets and financial portfolio implications}
   
\author[label1]{Nick James} \ead{nick.james@unimelb.edu.au}
\author[label2]{Kevin Chin} 
\address[label1]{School of Mathematics and Statistics, University of Melbourne, Victoria, Australia}
\address[label2]{Arowana \& Co.}

\begin{abstract}
This paper uses new and recently introduced mathematical techniques to undertake a data-driven study on the systemic nature of global inflation. We start by investigating country CPI inflation over the past 70 years. There, we highlight the systemic nature of global inflation with a judicious application of eigenvalue analysis and determine which countries exhibit most "centrality" with an inner-product based optimization method. We then turn to inflationary impacts on financial market securities, where we explore country equity indices' equity robustness and the varied performance of equity sectors during periods of significant inflationary pressure. Finally, we implement a time-varying portfolio optimization to determine which asset classes were most beneficial in increasing portfolio Sharpe ratio when an investor must hold a core (and constant) allocation to equities. 

\end{abstract}


\end{frontmatter}

\section{Introduction}
\label{Inflation_background}

Inflation has been dormant for the best part of 20 years in most developed countries globally, providing an environment for increasing business prosperity and consistent valuation multiple expansion across a number of asset classes including equities. In the wake of the COVID-19 pandemic and the associated economic and financial market crises, we are left with a new question of great pertinence - is the current elevation in global inflation levels transitory, or symptomatic of something more pathological? Rather than seeking to predict what is likely to happen next, we analyze previous periods of inflationary pressure and the associated impacts on economies, equity indices, equity sectors and investor portfolios. Our analysis takes a data-driven and mathematical approach, seeking to learn from the results of bespoke experiments, rather than approaching this problem with (often rigid) econometric theory.

Inflation is a well researched problem in the quantitative finance and economics literature \cite{Siu2011, Mishkin1990, Fama1990, Mundell1963}. A significant amount of prior work on the topic has applied well-established econometric theory  \cite{Tobin1965} to gain insights into the time-varying nature of inflationary behaviours. In the academic literature, however, there is limited work exploring the explicit impact of inflation on equity markets. Anecdotally, low-to-moderate inflation is typically associated with buoyant periods for equity markets. Inflationary extremes, where inflation may be particularly high or low, is generally considered to be harmful to equity market returns. In this paper, we test this hypothesis in a variety of contexts. Importantly, we study periods of time consistent with significant changes in inflation, rather than simply studying periods of high or low inflation. This way, we may ensure that our findings are relevant and topical in a period that is characterised by such pronounced economic uncertainty. 

More generally, however, the study of financial market dynamics has been a topic of great interest to the applied mathematics, econometrics and econophysics communities. In more recent years the study of time-varying dynamics and correlation structures \citep{Fenn2011,Laloux1999,Mnnix2012} has attracted the interest of researchers. A wide variety of techniques have been used to study these evolutionary dynamics including principal components analysis (PCA) \cite{Laloux1999, Kim2005, Pan2007, Wilcox2007}, clustering \cite{Heckens2020, Jamesfincovid}, change point detection \cite{James2021_crypto, JamescryptoEq} and various statistical modelling frameworks \cite{Tilfani2022, Ferreira2020}. Such analysis has been applied to a wide variety of underlying security types including equities \cite{Wilcox2007}, foreign exchange \cite{Ausloos2000}, and fixed income instruments \cite{Driessen2003}.

There has been a recent surge in interest among researchers studying various aspects of cryptocurrencies, the blockchain and other aspects of decentralised finance. Recent work that has attracted research interest include analyses of Bitcoin's behaviour \cite{Chu2015, Lahmiri2018, Kondor2014, Bariviera2017, AlvarezRamirez2018}, fractal-type behaviour and fractal patterns, \cite{Stosic2019, Stosic2019_2, Manavi2020, Ferreira2020}, cross-correlations and various scaling effects \cite{Drod2018, Drod2019, Drod2020, Gbarowski2019, Wtorek2020}, and trading strategies \cite{Fil2020}. Quite separately, the topic of portfolio optimization has been of significant interest to the quantitative finance community for many years \cite{Markowitz1952, Sharpe1966}, and the core problem has been extended upon considerably since its inception \cite{Almahdi2017, Calvo2014, Soleimani2009, Vercher2007, Bhansali2007, Moody2001}. Although there is literature examining the effectiveness of various asset classes' role as a portfolio `safe-haven', we are unaware of any work explicitly studying the benefit of cryptocurrencies in the context of extreme inflationary periods. This work studies the effect of cryptocurrencies in multi-asset portfolios, exploring their time-varying optimal portfolio allocation.

Many of these studies are concerned with the time-varying nature of such dynamics, or the behaviour of cryptocurrencies during various market regimes. Quite naturally, the impact of COVID-19 on cryptocurrency behaviours has been widely studied \cite{Corbet2020, Conlon2020, Conlon2020_2, Ji2020}. For recent work on time-varying similarity and its implications for the cryptocurrency market, and the significant impact of societal physics on the economy, readers should see \cite{Sigaki2019, Perc_social_physics, Perc2019}.

In Section \ref{Country_inflation_behaviours}, we examine the similarity in country CPI trajectories and the associated collective similarity in global inflation. We then introduce a new optimization framework to identify countries that are most "central" in global inflation, and which countries' equity indices are most robust during periods of significant inflationary change. Then in Section \ref{Equity_sector_dynamics} we turn to inflationary impacts on equity market performance, where we study the evolutionary correlation structure of various equity sectors. Finally in Section \ref{Dynamic_multi_asset_optimisation}, we implement a time-varying portfolio optimization over the past 5.5 years to determine which asset class would serve as the optimal complement to a core equity exposure, with the purpose of maximizing an investor's Sharpe ratio. Our methods demonstrate that, despite the associated volatility, cryptocurrency was the most essential additional asset class.

\section{Data}
\label{Data}

In Section \ref{Country_inflation_behaviours} we study country CPI data between 01/01/1955-01/09/2021 and equity index data 01/01/1990 - 01/09/2021. This data comes from a variety of sources including The International Monetary Fund (https://www.imf.org/en/Home), Bank of Canada (https://www.bankofcanada.ca/), Organisation for Economic Cooperation and Development (https://www.oecd.org/australia/), Bank of Japan (https://www.boj.or.jp/en/index.htm/) and Bloomberg. In Section \ref{Equity_sector_dynamics} we study equity sector behaviours based on underlying equity data between 01/01/2005 - 31/12/2009, this data is sourced from Bloomberg. In Section \ref{Dynamic_multi_asset_optimisation} we study various asset class performance (in the context of portfolio optimization) between 01/01/2016-30/06/2021. This data is sourced from Bloomberg and Coinmarketcap.com (https://coinmarketcap.com/).

\section{Evolutionary country inflation dynamics}
\label{Country_inflation_behaviours}

\subsection{CPI Inflation trajectories}
\label{inflation_trajectories}
In this section of the paper, we study a collection of $n=8$ countries listed alphabetically and indexed $i=1,...,n$. The countries in this section include: Australia, Canada, France, Germany, Italy, Japan, the United Kingdom (UK) and the United States of America (USA). We study monthly CPI inflation data between 01/01/1955 to 01/09/2021, a total of $T=801$ months. Let $x_i(t) \in \mathbb{R}$ be the multivariate time series of CPI inflation for country $i$ in month $t$. We wish to study the evolutionary change in inflation over time, and accordingly we compute the log returns of each country's CPI inflation and contrast these behaviours over time. Each country's CPI log returns are computed as follows

\begin{align}
\Delta^{x}_{i}{(t)} &= \log \left(\frac{x_i{(t)}}{x_i{(t-1)}}\right),
\end{align}

while $\Delta^{x}_{i}{(t)}$, references country $i$'s CPI log returns at time $t$. To compare inflation trajectories of countries where raw inflation varies significantly, the CPI returns time series are normalized. Each country generates an inflation returns time series, $\mathbf{\Delta}^{x}_{i} \in \mathbb{R}^{T}$. We let $\|\mathbf{\Delta}^{x}_{i}\|_1 = \sum^{T}_{t=1} |\Delta^{x}_{i}(t)|$ be the $L^1$ norm of the inflation returns function, and define a normalized inflation returns trajectory by $\mathbf{T}^{x}_{i} = \frac{\mathbf{\Delta}^{x}_i}{\|\mathbf{\Delta}^{x}_i\|_1}$. Any distance between two countries' trajectory vectors would highlight the relative similarity in CPI changes over time. We compute an \emph{inflationary trajectory matrix} $D^{x}_{ij}=\|\mathbf{T}^x_i - \mathbf{T}^x_j\|_1$, and apply \emph{hierarchical clustering} to the resulting distance matrix.

\begin{figure}
    \centering
    \begin{subfigure}[b]{0.48\textwidth}
        \includegraphics[width=\textwidth]{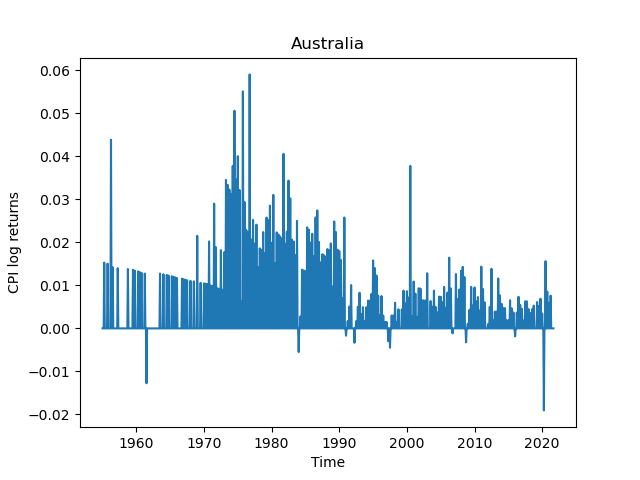}
        \caption{}
        \label{fig:CPI_Australia}
    \end{subfigure}
    \begin{subfigure}[b]{0.48\textwidth}
        \includegraphics[width=\textwidth]{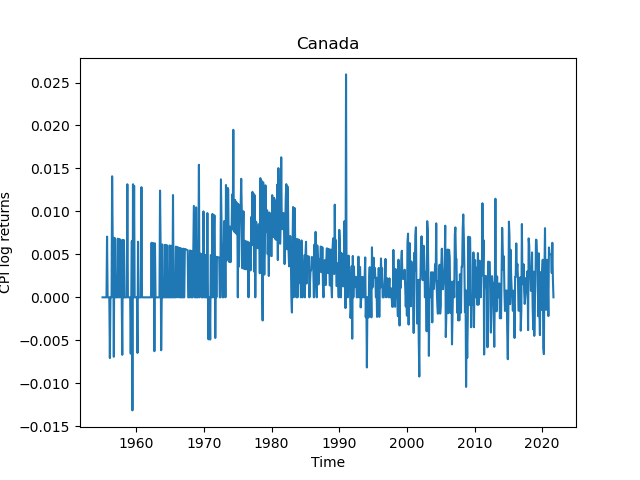}
        \caption{}
        \label{fig:CPI_Canada}
    \end{subfigure}
    \begin{subfigure}[b]{0.48\textwidth}
        \includegraphics[width=\textwidth]{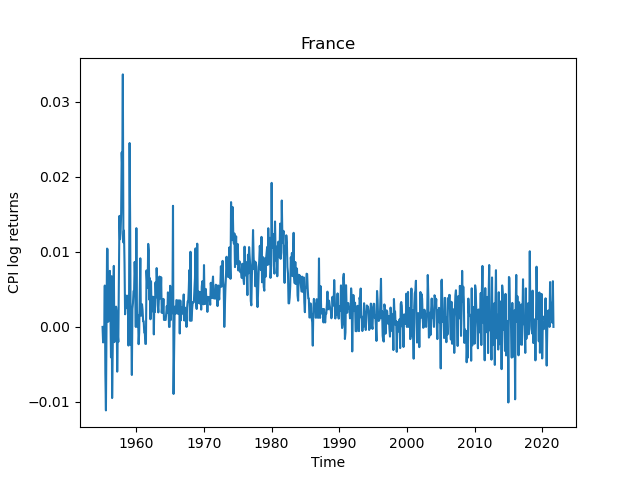}
        \caption{}
        \label{fig:CPI_France}
    \end{subfigure}
    \begin{subfigure}[b]{0.48\textwidth}
        \includegraphics[width=\textwidth]{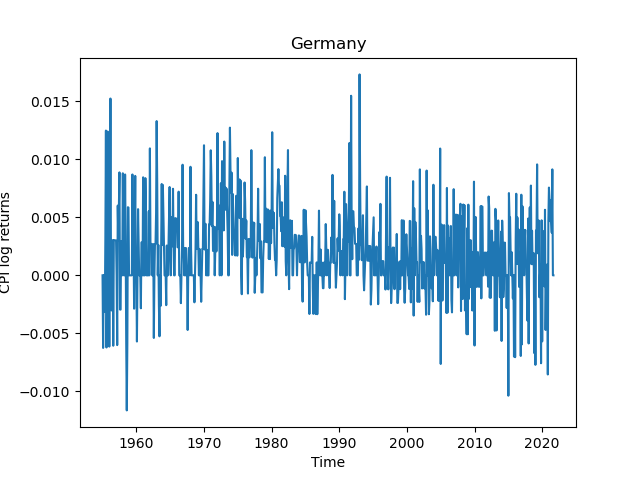}
        \caption{}
        \label{fig:CPI_Germany}
    \end{subfigure}
    \begin{subfigure}[b]{0.48\textwidth}
        \includegraphics[width=\textwidth]{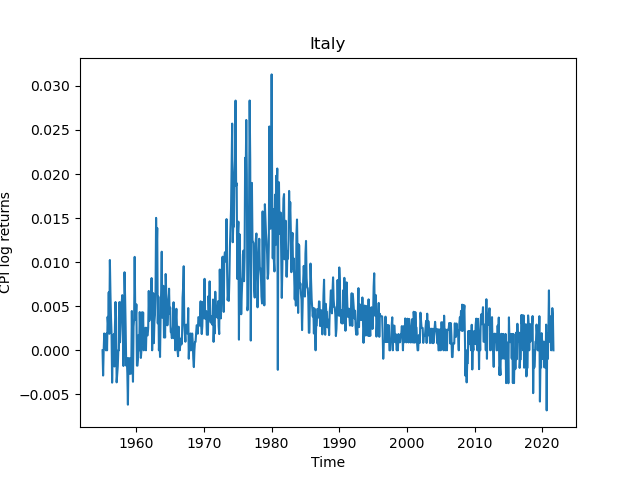}
        \caption{}
        \label{fig:CPI_Italy}
    \end{subfigure}
    \begin{subfigure}[b]{0.48\textwidth}
        \includegraphics[width=\textwidth]{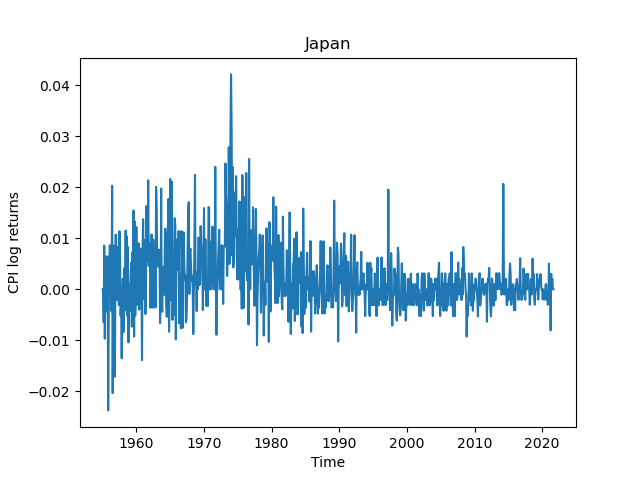}
        \caption{}
        \label{fig:CPI_Japan}
    \end{subfigure}
        \begin{subfigure}[b]{0.48\textwidth}
        \includegraphics[width=\textwidth]{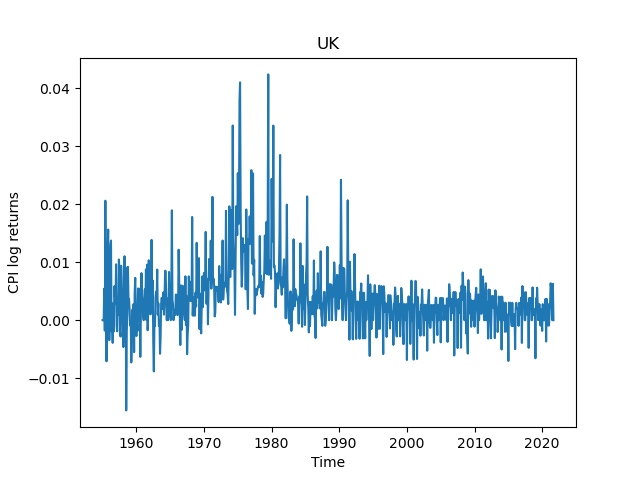}
        \caption{}
        \label{fig:CPI_UK}
    \end{subfigure}
    \begin{subfigure}[b]{0.48\textwidth}
        \includegraphics[width=\textwidth]{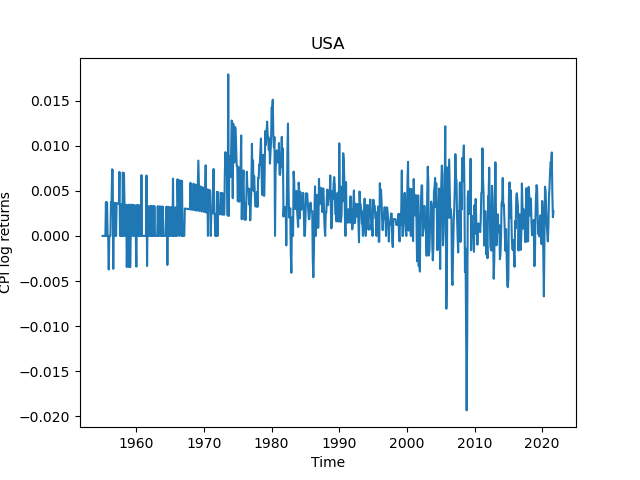}
        \caption{}
        \label{fig:CPI_USA}
    \end{subfigure}
    \caption{CPI log return time series (a) Australia (b) Canada, (c) France, (d) Germany, (e) Italy, (f) Japan, (g) UK, (h) USA. There are some notable similarities in the evolutionary inflation differences including spikes around 1980. Australia and Japan appear to exhibit a spike in inflationary differences slightly earlier than other countries.}
    \label{fig:CPI_time_series}
\end{figure}

Figure \ref{fig:CPI_time_series} reveals several noticeable similarities between various countries' inflation between 1955-2021. The most prominent similarity is the inflationary spike exhibited by all countries during the 1970s and early 1980s. Most countries (including Canada, France, Germany, Italy, UK and the USA shown in Figures \ref{fig:CPI_Canada}, \ref{fig:CPI_France}, \ref{fig:CPI_Germany}, \ref{fig:CPI_Italy}, \ref{fig:CPI_UK} and \ref{fig:CPI_USA} respectively) display similarly-timed peaks in inflationary changes. The two notable outliers in this regard are Australia and Japan (Figures \ref{fig:CPI_Australia} and \ref{fig:CPI_Japan}) who exhibit earlier inflationary spikes than other countries. The second theme that is broadly consistent among countries are sharp drops in inflation corresponding to the GFC and COVID-19 market crises. Although there is some variability among all countries, the most interesting trajectory is that of the USA which exhibits the most pronounced shock during the GFC, followed by relatively buoyant inflationary levels. Other countries such as France, Germany, Italy and the UK exhibit negative inflationary differences more frequently than other countries. 


\begin{figure}
    \centering
    \includegraphics[width=0.95\textwidth]{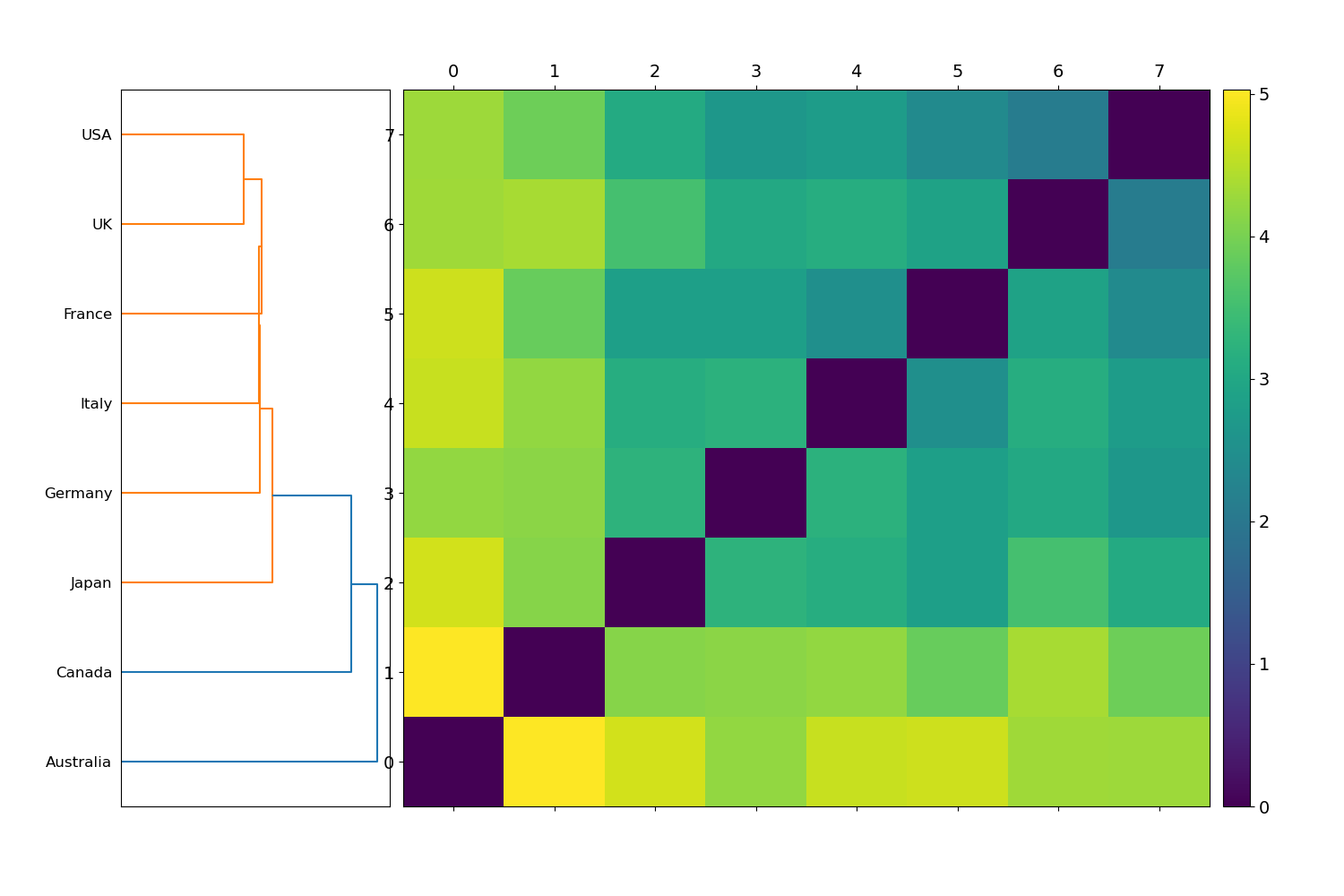}
    \caption{Hierarchical clustering applied to the distance matrix $D^{x}$. One predominant country cluster is formed with two outliers detected - Australia and Japan. This is most likely due to the earlier onset of their inflationary pressures. }
    \label{fig:CPI_dendrogram}
\end{figure}

To further investigate overall country inflationary structure, we apply hierarchical clustering to our inflationary trajectory matrix $D^{x}$. dendrogram displayed in Figure \ref{fig:CPI_dendrogram} reveals a cluster structure consistent with the CPI log return time series shown in Figure \ref{fig:CPI_time_series}. The dendrogram reveals one predominant cluster consisting of the USA, Canada, France, Italy, Germany and the UK. The two outlier countries identified are Japan and Australia. Their separate classification with respect to the remaining countries is most likely due to the spike in inflation levels which precedes the bulk of countries.

\subsection{Collective country inflation behaviours}
\label{Collective_country_inflation_behaviours}

Having identified two outlier countries with our trajectory analysis, we further explore the structural similarity between countries' CPI inflationary patterns with a judicious application of eigenvalue analysis. Given the results from our methods in preceding sections, we expect the majority of inflation time series to display significant similarity over time. 
     
We start with the following contrived scenario. Consider a large collection of financial time series, $50$, for example. If $40$ out of the $50$ time series display a high degree of affinity, then the distance matrix $D^{x}$ should have the following structure:

\[
\begin{pmatrix}
  \begin{matrix}
   & c_{1} & c_{2} & c_{3} & \hdots & c_{40} \\
  r_{1} & 0 & * & * & \hdots & * \\
  r_{2} & * & 0 & * & \hdots & * \\
  r_{3} & * & * & 0 & \hdots & *\\
   & \vdots & \vdots & \vdots & \ddots \\
  r_{40} & * & * & * & * & 0 \\
  \end{matrix}
  & \rvline &   
  \begin{array}{ccc}
    \horzbar & r_{1} & \horzbar \\
    \horzbar & r_{2} & \horzbar \\
             & \vdots    &          \\
    \horzbar & r_{40} & \horzbar
  \end{array} & \\
\hline
  \begin{array}{ccccc}
    \vertbar & \vertbar & \vertbar &        & \vertbar \\
    r^{T}_{1}    & r^{T}_{2} & r^{T}_{3}   & \ldots & r^{T}_{40}    \\
    \vertbar & \vertbar & \vertbar &       & \vertbar 
  \end{array} & \rvline &
  \begin{matrix}
  0 &  \\
   & 0 \\
   &  & \ddots \\
   & & & 0
  \end{matrix}
\end{pmatrix}
\]
where rows $r_1,\ldots,r_{40}$ are time series that display a high degree of affinity and elements $*$ are close to zero. This interpretation of this is that small deformations in elements of the matrix make the first $40$ rows identical. Therefore, the matrix is a small perturbation from a rank $11$ matrix, with $39$ eigenvalues that are equal to zero. So in this hypothetical case, if $40$ out of our $50$ time series exhibit similar behaviours, then $39$ of our eigenvalues would be close to zero.

We now return to our current data. Given a candidate threshold $\delta$, we are able to order the absolute values of the distance matrix eigenvalues $|\lambda_1|\leq...\leq|\lambda_n|$. If $|\lambda_1|,...,|\lambda_k|<\delta$ then we can deduce $k+1$ of our country CPI inflation time series are similar with respect to their evolutionary behaviours. Eigenvalue analysis such as this provide a fast and intuitive way of assessing the scale of any distance matrix. Since our distance matrix is symmetric, it can be conjugated by an orthogonal matrix which yields a diagonal matrix of eigenvalues (this is also known as the spectral theorem). The result is that the operator norm of the matrix coincides with $|\lambda_n|$, or, $$\max_{y \in \mathbb{R}^n} \frac{||D^{x}y||}{||y||}= ||D^{x}||_{oper}= |\lambda_n|$$.

\begin{figure}
    \centering
    \includegraphics[width=0.95\textwidth]{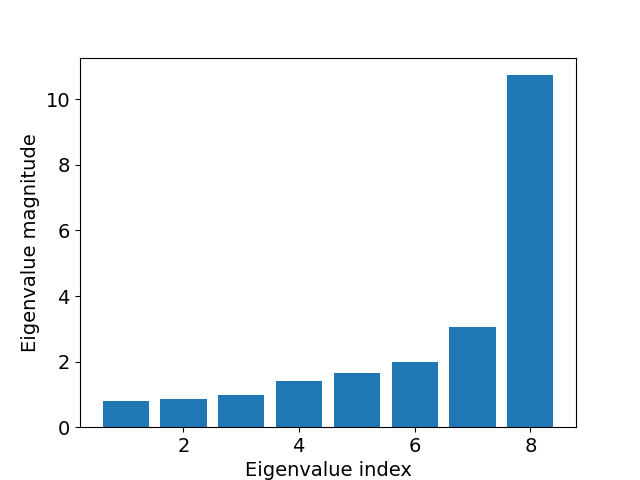}
    \caption{Eigenspectrum of the distance matrix $D^{x}$. This is perhaps the most concise and expressive way of detailing the collective similarity in evolutionary inflation patterns. Eigenvalue analysis suggests that there are 1 or 2 countries that are dissimilar to the rest of the collection. }
    \label{fig:Inflation_eigenspectrum}
\end{figure}

Figure \ref{fig:Inflation_eigenspectrum} shows the magnitude of eigenvalues associated with the distance matrix, $D^x$. It is clear from the eigenspectrum that there is a large collection of countries with highly similar CPI inflation trajectories. For our purposes, we set $\delta = 2.5$, and identify two outlier countries (consistent with both findings in the preceding section \ref{inflation_trajectories}. Again, Australia and Japan are identified as the most dissimilar countries based on our eigenvalue analysis. 


\subsection{Inner product-based "Centrality-identification"}
\label{lead_lag_inflation_coefficients}

Given the predominant structural similarity in country CPI inflation that we have now identified, we wish to further study the systemic nature of such inflation and determine whether some countries are more `central' to global inflation than others. To examine this phenomenon, we implement a time-varying linear regression on inflation log returns, and study the evolution of each model's coefficients. Individually, this methodology allows us to see the evolution of inflationary trends over time. When viewed in conjunction with other countries' results, this framework provides us with a general overview of the time-varying global similarity in inflationary trends. To explore the dynamic nature of each country's CPI inflation, we study the evolution of model coefficients in two settings. Model 1 uses a 60-month rolling window, and model 2 uses a 120-month rolling window. In each case, we restrict our model to specific intervals of either 60 months or 120 months, and study the evolution of of model coefficients over time. We now compute:

\begin{align}
   \Delta^{x}_i [t-60:t]  = \beta_{0,i} [t-60:t]  + \beta_{1,i} [t-60:t]  +  \epsilon_{i} [t-60:t], \text{ and} \\
   \Delta^{x}_i [t-120:t]  = \beta_{0,i} [t-120:t] + \beta_{1,i} [t-120:t]  +  \epsilon_{i} [t-120:t],
\end{align}

for Model 1 and Model 2 respectively. Implementing our model for all $i$ countries in the data yields a vector of \emph{evolutionary trend coefficients} which we label $\mathbb{\beta}^{M_1}_{1,i}(t) \forall t \in \{61,...,T\}$ and $\mathbb{\beta}^{M_2}_{1,i}(t)  \forall t \in \{121,...,T\}$, for models 1 and 2 respectively. The choice of our smoothing windows is quite deliberate. For the 60-month rolling window, we wish to study inflationary trends in a more dynamic manner, using 5 year windows. For the study of longer term country inflationary patterns, we use a 120-month rolling window, as is reflected in Model 2. Although there is literature to suggest that macroeconomic cycles typically last somewhere in the order of 5-7 years, we do not wish for excessive transience in our regression parameters. 

Although we study an evolutionary trend coefficient with a time-varying linear regression implementation, other modelling approaches could be used. Some possibilities of alternative models include moving average (MA) and autoregressive moving average (ARMA) models. Such analysis could be the foundation of future research, where one could study the capacity for change in the two polynomial terms among country CPI time series. Furthermore, one could explore the existence of change points and structural breaks based on changes in the behaviour of the autoregressive model parameters.

\begin{figure}
    \centering
    \begin{subfigure}[b]{0.95\textwidth}
        \includegraphics[width=\textwidth]{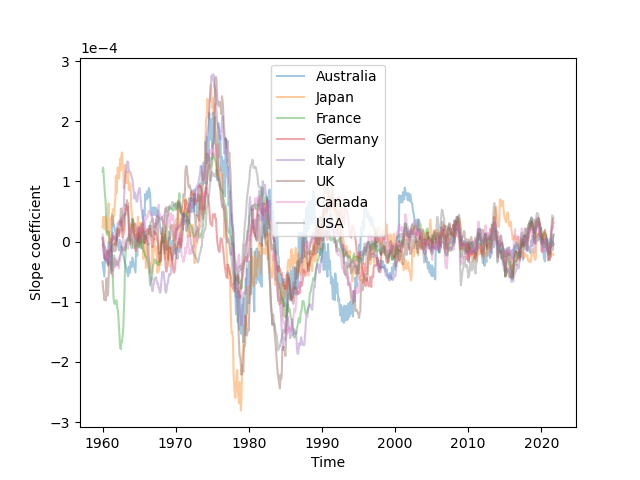}
        \caption{}
        \label{fig:Slope_coefficient_60}
    \end{subfigure}
    \begin{subfigure}[b]{0.95\textwidth}
        \includegraphics[width=\textwidth]{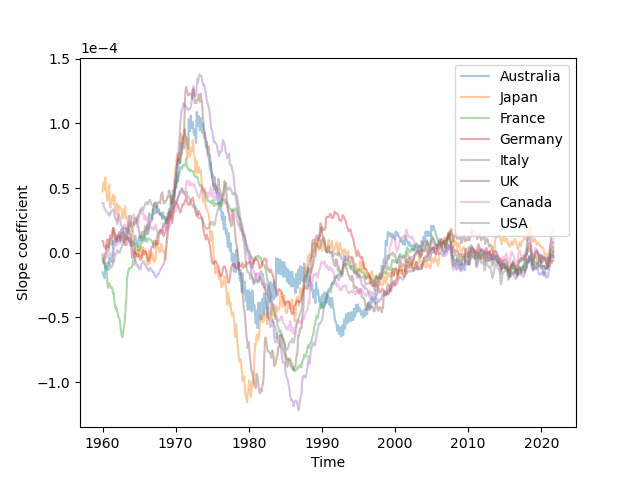}
        \caption{}
        \label{fig:Slope_coefficient_120}
    \end{subfigure}
    \caption{Evolutionary slope coefficients for all 8 countries studied for (a) Model 1 (60-month rolling window) and (b) Model 2 (120-month rolling window). Most countries exhibit a broadly similar trend of "up-down-up" early in their trajectories (corresponding to the 1970s, 1980s and 1990s, respectively) This is followed by a prolonged period of slope coefficients of smaller magnitude.}
    \label{fig:Slope_coefficients}
\end{figure}

The general findings for both models are relatively consistent. Both Model 1 and Model 2, shown in Figures \ref{fig:Slope_coefficient_60} and \ref{fig:Slope_coefficient_120} respectively display a general "up-down-up" pattern of inflation trends. That is they display a positive inflation trend during the 1970s, followed by a negative trend in the 1980s and a reversion to normal levels by the early 2000s. From the early 2000s onwards, all country slope coefficients are of significantly lower magnitude, and display greater affinity between each other. One observation that can be made from both figures, is that although the general shape of inflationary trends are highly similar between countries, some countries appear to precede others in reaching local maxima and minima in slope coefficient magnitudes. This raises several interesting questions. First, which countries display the greatest differences to the rest of the collection with respect to their inflation behaviours? Second, which countries are most "central" in systemic global inflation. 


To explore these questions, we introduce an inner product-based optimisation framework to determine a suitable translation (between all candidate country pairings) that would align country inflation trends most appropriately. For each country, we seek to learn an offset $\phi$, such that we maximize a normalized inner product. In the proceeding section where we describe our optimization methodology, we drop the superscripts $M_1$ and $M_2$ denoting Models 1 and 2, for the sake of more accessible exposition.

For any two countries that we compare, we allow $\phi$ to be either positive or negative, so the general approach is flexible to the possibility that any country inflation time series may be leading or lagging another. We now define our normalized inner product between any two evolutionary trend coefficient trajectories. For any two countries' slope trajectories $\beta_{1,i}$ and $\beta_{1,j}$ we define two possible inner product expressions, based on the direction of our offset $\phi$. If $\phi > 0$, we may express our objective function
\begin{align}
<\beta_{1,i}(0:T-\phi),\beta_{1,j}(\phi:T)>_n =
\frac{\beta_{1,i}(0)\beta_{1,j}(\phi)+...+\beta_{1,i}(T-\phi)\beta_{1,j}(T)}{(\beta_{1,i}(0)^2+...+\beta_{1,i}(T-\phi)^2)^\frac{1}{2}(\beta_{1,j}(\phi)^2+...+\beta_{1,j}(T)^2)^\frac{1}{2}},
\end{align}


If $\phi<0$ we let $\sigma = - \phi$, and this can now be expressed
\begin{align}
<\beta_{1,i}(\sigma:T),\beta_{1,j}(0:T-\sigma)>_n =
\frac{\beta_{1,i}(\sigma)\beta_{1,j}(0)+...+\beta_{1,i}(T)\beta_{1,j}(T-\sigma)}{(\beta_{1,i}(\sigma)^2+...+\beta_{1,i}(T)^2)^\frac{1}{2}(\beta_{1,j}(0)^2+...+\beta_{1,j}(T-\sigma)^2)^\frac{1}{2}}.
\end{align}

We record the optimal offset between all country slope trajectories in an \emph{inflation offset matrix} which we denote $D^{M_1}_{\phi}$ and $D^{M_2}_{\phi}$, for Models 1 and 2 respectively. To determine countries that are most and least central in the evolution of global inflation, we compute an \emph{inflation centrality score} for each matrix as follows $centrality^{M_1}_j = \sum_{i} D^{M_1}_{\phi, ij}$, and $centrality^{M_2}_j = \sum_{i} D^{M_2}_{\phi, ij}$. The interpretation of this measure is that the lower the centrality score, the closer any country is to the rest of the collection.

\begin{table}
\centering
\begin{tabular}{ |p{2.9cm}||p{2cm}|p{2cm}|}
 \hline
 \multicolumn{3}{|c|}{Country inflation centrality scores} \\
 \hline
 Country & $centrality^{M_1}$ & $centrality^{M_2}$ \\
 \hline
 Australia & 144 & 94 \\
 Canada & 38 & 49 \\
 France & 74 & 146 \\
 Germany & 78 & 86 \\
 Italy & 14 & 64 \\
 Japan & 130 & 199 \\
 UK & 43 & 54 \\
 USA & 81 & 56 \\
\hline
\end{tabular}
\caption{Inflation centrality scores for all countries under Model 1 and Model 2. In both cases, there is a majority collection of central countries. Under both models, Japan is most consistently the least central (most anomalous) country.}
\label{tab:Inflation_centrality_scores}
\end{table}

\begin{figure}
    \centering
    \begin{subfigure}[b]{0.75\textwidth}
        \includegraphics[width=\textwidth]{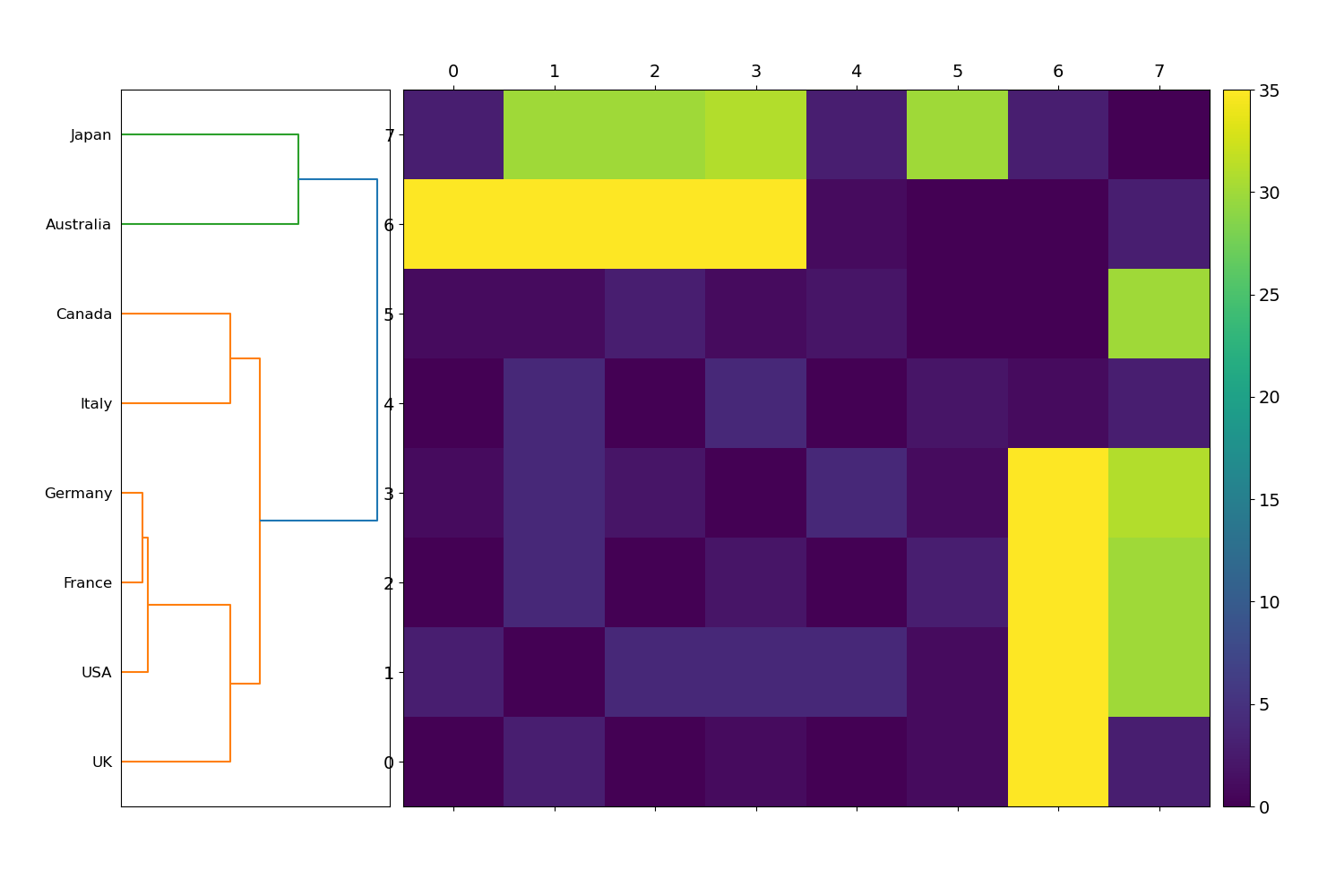}
        \caption{}
        \label{fig:Slope_coefficient_dendrogram_60}
    \end{subfigure}
    \begin{subfigure}[b]{0.75\textwidth}
        \includegraphics[width=\textwidth]{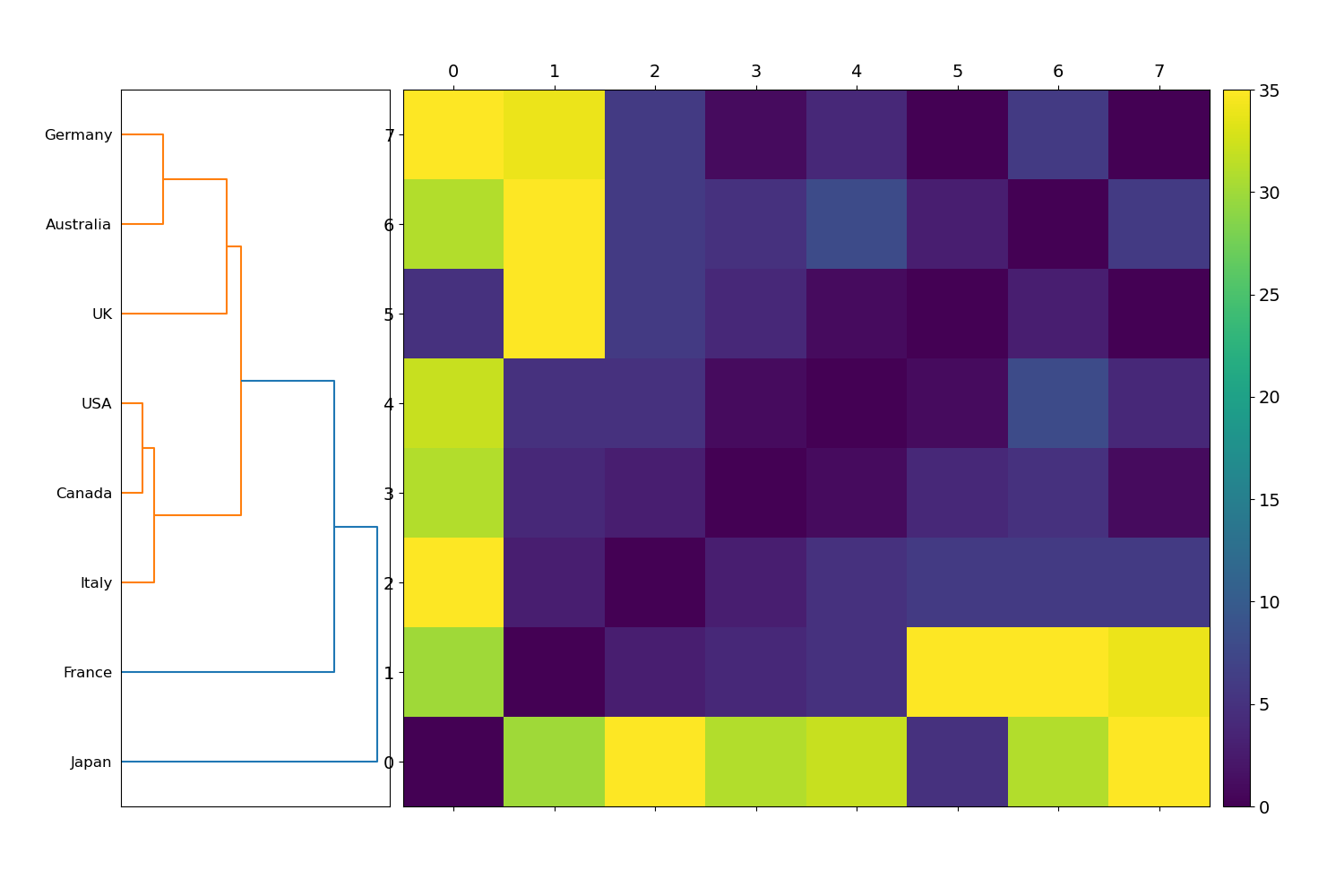}
        \caption{}
        \label{fig:Slope_coefficient_dendrogram_120}
    \end{subfigure}
    \caption{Hierarchical clustering applied to the two distance matrices (a) $D^{M_1}_{\phi}$ and (b) $D^{M_2}_{\phi}$. The dendrogram shows that under both models, there is a majority cluster of countries who display similar timing in global inflationary trends. Japan is most consistently the "least central" country with respect to its inflation behaviours.}
    \label{fig:Slope_coefficients_dendrogram}
\end{figure}

We further investigate the evolution of each country's inflation centrality by partitioning our analysis into two discrete windows. First, we study inflation centrality between January 1960 to December 1989, and then study the period between January 1990 and December 2019. It is clear from Figure \ref{fig:Slope_coefficients} that the period until 1990 exhibits significantly higher inflation - and greater volatility in country inflation trends. The period between 1990 and 2020 exhibits significantly lower inflation, and less variability in country inflation trends. Given that we have halved the length of our analysis window, we introduce a new model (Model 3) and use a 30-month rolling window. 

\begin{table}
\centering
\begin{tabular}{ |p{2.8cm}||p{2.6cm}|p{2.6cm}|}
 \hline
 \multicolumn{3}{|c|}{Country inflation centrality scores: partitioned analysis} \\
 \hline
 Country & $centrality^{M_3}$: 1960-1990 & $centrality^{M_3}$: 1990:2020 \\
 \hline
 Australia & 31 & 47 \\
 Canada & 14 & 22  \\
 France & 23 & 13  \\
 Germany & 35 & 66  \\
 Italy & 10 & 22  \\
 Japan & 14 & 73  \\
 UK & 24 & 44  \\
 USA & 23 & 3  \\
\hline
\end{tabular}
\caption{Inflation centrality scores for all countries under Model 3 (30-month rolling window) between 1960-1990 and 1990-2020.}
\label{tab:Inflation_centrality_scores_partition}
\end{table}

Table \ref{tab:Inflation_centrality_scores_partition} captures some interesting insights. In our first partition, studying the period between January 1960 to December 1989, there is a material difference in inflation centrality structure. Most significantly, Australia and Germany are identified to be most anomalous (with respective scores of 31 and 35), while Japan is much more central in its inflationary behaviours. During our second analysis window which studies the period between January 1990 and December 2019, Japan is identified as the most anomalous (least central) country with respect to its inflationary behaviours. It is worth noting that Japan's identification as an inflationary anomaly is most pronounced during a period of relative stability in global inflation.

\subsection{Country equity index robustness}
\label{Country_equity_index_robustness}

For our 8 countries examined in previous sections, we now wish to study the robustness of each country's equity market during extreme inflationary periods. To do so, we introduce a new multivariate time series $y_i(t)$, which refers to each country's respective equity index. We study data between 01/01/1990 - 01/09/2021, a total of $T_1 = 381$ months. We index this window of time $t_1=1,...,T_1$.

We compute the log returns of each equity index as follows 

\begin{align}
\Delta^{y}_{i}{(t_1)} &= \log \left(\frac{y_i{(t_1)}}{y_i{(t_1-1)}}\right).
\end{align}

From each country's log returns time series, we construct two truncated \emph{probability density functions} (PDFs). The first consists of the country equity index returns during the top decile and bottom decile of inflationary returns. That is, we extract equity index returns consistent with a country's most extreme inflationary and deflationary pressures. The second distribution consists of equity index returns during the remaining months. Let the first pdf containing extreme equity returns be $g^{ex}_i(x)$, and the stable period pdf be $g^{st}_i(x)$. For each country $i$, we compute the distance between the two pdfs. A small distance would indicate less extreme equity behaviours during inflationary crises, while a large distance may show countries that are more prone to pronounced equity market fluctuations during strongly inflationary and deflationary periods. For each country, we measure the distance between these two distributions with a \emph{Wasserstein distance}\cite{Kantorovich1960}, and term the resulting measure an \emph{equity robustness} distance

\begin{align}
ER_i = d^{W} (g^{st}_i(x), g_i^{ex}(x)),
\end{align}

\begin{figure}
    \centering
    \begin{subfigure}[b]{0.48\textwidth}
        \includegraphics[width=\textwidth]{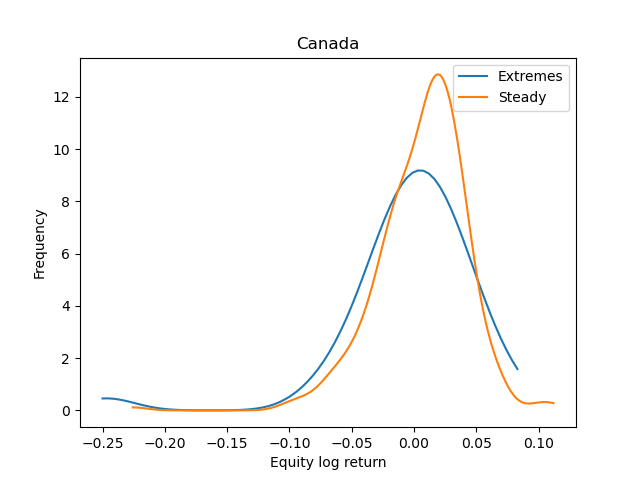}
        \caption{}
        \label{fig:Distribution_Canada}
    \end{subfigure}
    \begin{subfigure}[b]{0.48\textwidth}
        \includegraphics[width=\textwidth]{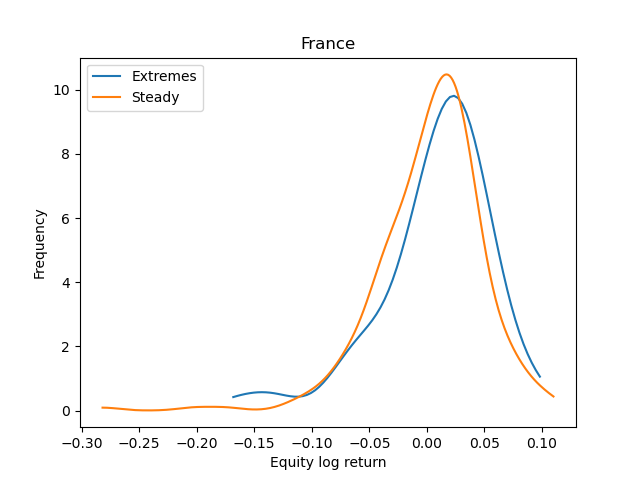}
        \caption{}
        \label{fig:Distribution_France}
    \end{subfigure}
    \begin{subfigure}[b]{0.48\textwidth}
        \includegraphics[width=\textwidth]{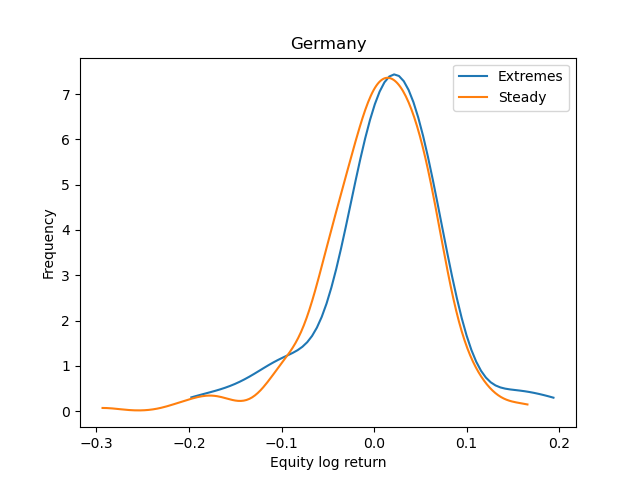}
        \caption{}
        \label{fig:Distribution_Germany}
    \end{subfigure}
    \begin{subfigure}[b]{0.48\textwidth}
        \includegraphics[width=\textwidth]{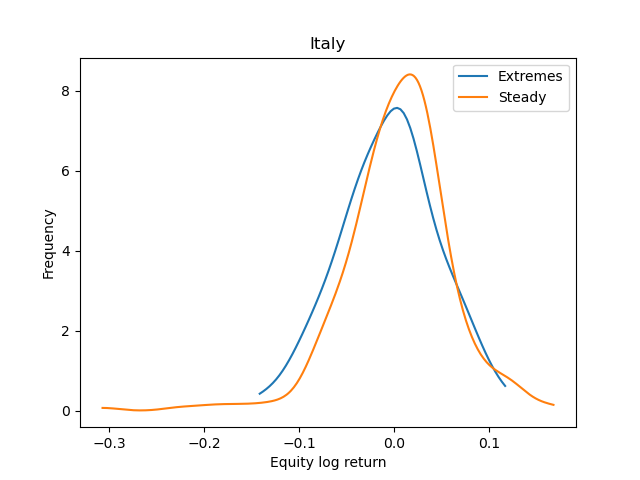}
        \caption{}
        \label{fig:Distributions_Italy}
    \end{subfigure}
    \begin{subfigure}[b]{0.48\textwidth}
        \includegraphics[width=\textwidth]{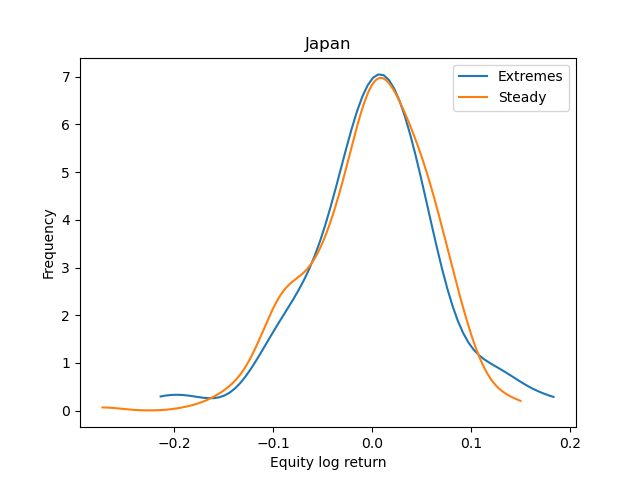}
        \caption{}
        \label{fig:Distributions_Japan}
    \end{subfigure}
        \begin{subfigure}[b]{0.48\textwidth}
        \includegraphics[width=\textwidth]{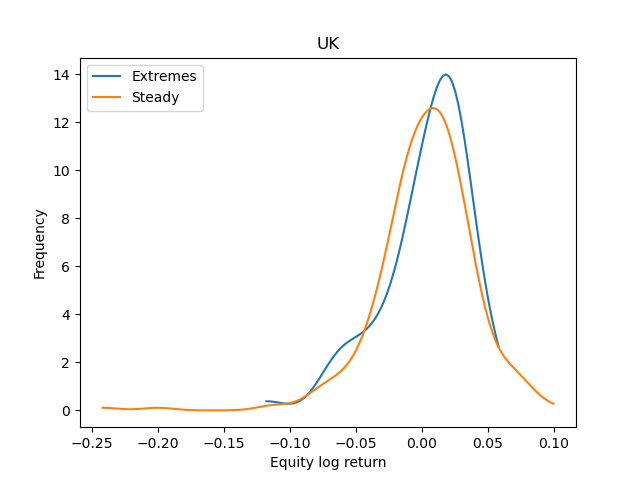}
        \caption{}
        \label{fig:Distributions_UK}
    \end{subfigure}
    \begin{subfigure}[b]{0.48\textwidth}
        \includegraphics[width=\textwidth]{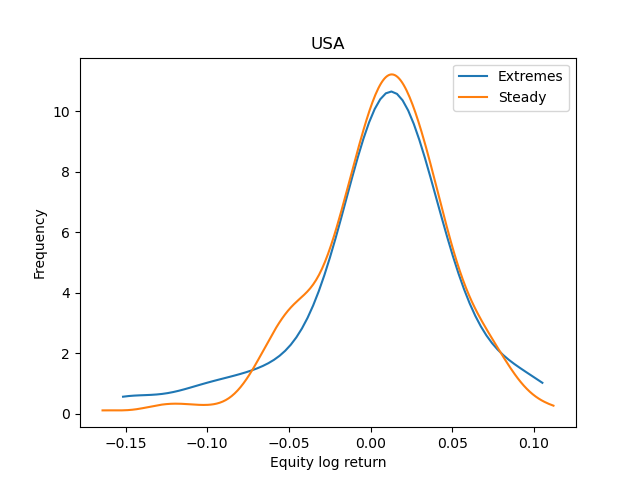}
        \caption{}
        \label{fig:Distributions_USA}
    \end{subfigure}
    \caption{Country equity distributions during extreme and stable inflationary periods for (a) Canada, (b) France, (c) Germany, (d) Italy, (e) Japan, (f) UK, (g) USA. For the purposes of visualisation, we have applied kernel density estimation to all countries' two conditional distributions. Australian data is only available on a sparser periodicity, and Australia has been excluded for this reason.}
    \label{fig:Distributions_time_series}
\end{figure}

\begin{table}
\centering
\begin{tabular}{ |p{2.9cm}||p{2.5cm}|}
 \hline
 \multicolumn{2}{|c|}{Country equity robustness} \\
 \hline
 Country & $ER$ \\
 \hline
 Canada & .012 \\
 France & .009 \\
 Germany & .010 \\
 Italy & .012 \\
 Japan & .008 \\
 UK & .006 \\
 USA & .006 \\
\hline
\end{tabular}
\caption{Country equity robustness scores computed using Wasserstein distance between two conditional distributions of log returns. The USA and the UK have equity indices that are most robust during extreme inflationary periods.}
\label{tab:Country_equity_robustness}
\end{table}

where $d^{W}$ is a Wasserstein distance between our two distributions. Results are shown in Table \ref{tab:Country_equity_robustness}. The table reveals that the countries exhibiting the smallest distance between the two distributions are the USA and UK. This implies that these countries' equity indices exhibit the greatest robustness to inflationary pressures. By contrast, Italy and Canada display the largest Wasserstein distance between the extreme and stable periods of inflationary market dynamics. This suggests that these countries' equity indices may display the least robustness during periods of extreme inflationary pressure.

As a subsequent experiment, we explicitly compare the difference in equity returns in high and low inflationary periods for all countries. The results highlight that the country with the smallest discrepancy in equity returns is the United States, with a difference in average monthly log returns of -0.003 between the respective inflationary periods (high and low). The country with the largest discrepancy in average log returns is Canada, which has a difference in average monthly log returns of 0.029 in periods of extremely high and low inflation. This suggests that the United States' equity index exhibits the most similar behaviours during high and low inflationary periods. By contrast, Canada exhibits the greatest difference in equity returns during high and low inflationary periods. Examining the rest of the collection, there is significant variability as to whether average log returns perform better in anomalously high or low inflationary periods. In Australia, Japan, Germany, the UK and Canada, equity indices outperform during extremely high inflationary periods (relative to extremely low inflationary periods), while in France, Italy and the USA the opposite occurs. Further research could test this phenomenon more explicitly, on a larger collection of countries and their respective equity indices.

\section{Underlying equity sector dynamics}
\label{Equity_sector_dynamics}

Having explored equity index robustness, we now focus on the performance of underlying equities, grouped by sector, during a more recent period of time. Here, we study correlation behaviours of S\&P 500 equities to determine differences in sector  behaviours during periods of significant inflationary pressure. We partition these equities into $S=11$ sectors, and study the evolutionary correlation dynamics of each sector. Each underlying equity is grouped into one sector from Communication Services, Consumer Discretionary, Consumer Staples, Energy, Financials, Healthcare, Industrials, Information Technology, Materials, Real Estate and Utilities. We study the evolution of intra-sector correlation dynamics over a relatively recent period that exhibited extreme inflationary behaviours, 1/1/2005-31/12/2009. Let $t_2=1,...,T_2$ be an index in time reflecting this date range. Let each sector's collection of underlying equity log returns be the vector $\mathbf{R}^{S}(t_2)$, where $S$ indexes one of 11 sectors, and $t_2$ denotes a specific point in time for the vector of log return values. We use a rolling window of 120 days. For each sector, we implement a time-varying correlation matrix as follows: 

\begin{align}
\Omega_{ij}^{S} (t_2) = \frac{\sum_{k=t_2-120}^{t_2} (R^{S}_i(k) - \bar{R}^{S}_i)(R^{S}_j(k) - \bar{R}^{S}_j))}{\left(\sum_{k=t_2-120}^{t_2} (R^{S}_i(k) - \bar{R}^{S}_i)^2 \sum_{k=t_2-120}^{t_2} (R^{S}_j(k) - \bar{R}^{S}_j)^2\right)^{1/2}}, \forall {t_2} \in \{121,...,T_2 \}
\end{align}

and study the evolution of each sector's correlation behaviours. For each evolutionary correlation matrix $\Omega_{ij}^{S} (t)$, we wish to compute an average sector correlation $\tilde{\mu}^{S}$ during our window of interest. To determine the average correlation of each sector during the period 1/1/2005-31/12/2009, we compute 

\begin{align}
\tilde{\mu}^{S} = \frac{2}{S_n (S_n - 1) \times 1303} \sum^{2610}_{t_2= 1307} \sum_{1\leq i<j\leq N} \Omega^{S}_{ij} (t_2) 
\end{align}

where $S_n$ is the number of equities in each sector, $t_2=1307$ and $t_2=2610$ correspond to 1/1/2005 and 31/12/2009 respectively. 

\begin{table}
\centering
\begin{tabular}{ |p{3.8cm}||p{2.9cm}|}
 \hline
 \multicolumn{2}{|c|}{Average sector correlation: 1/1/2005: 31/12/2009} \\
 \hline
 Country & $\tilde{\mu}^{S}$ \\
 \hline
 Communication Services & 0.40 \\
 Consumer Discretionary & 0.42 \\
 Consumer Staples & 0.34 \\
 Energy & 0.70 \\
 Financials & 0.50 \\
 Healthcare & 0.32 \\
 Industrials & 0.45 \\
 Information Technology & 0.42 \\
 Materials & 0.50 \\
 Real estate & 0.66 \\
 Utilities & 0.62 \\
\hline
\end{tabular}
\caption{Average sector correlation during the period 1/1/2005 - 31/12/2009. There is significant variability in the correlation of underlying equities based on the sector. The sectors with highest and lowest average correlation among constituent equities are energy and healthcare.}
\label{tab:Average_sector_correlation}
\end{table}

There are several reasons why we elect to study this period of time. First, rather than simply studying periods of high inflation, we wish to study equity performance during periods of significant changes in inflation (we are indifferent to the direction of such changes). In the current economic climate, the presence of inflation may be transient - and we could see a sharp spike in CPI followed by a reversion to more stable levels. Alternatively, we could see a structural increase in global inflation, and these levels may remain elevated for a prolonged period of time (such as that seen across the world in the 1970s and 1980s). Given the vast range of possibilities, we deliberately choose a country (USA) which experienced profound inflationary changes (both up and down) for a substantial period of time. Second, given the wealth of market commentary currently speculating on an imminent market crisis, we wish to study the behaviour of various equity sectors during a more prolonged crash such as the Global Financial Crisis (GFC). We avoid examining equity behaviours during the COVID-19 market crisis due to its brevity and the immediate nature of this market shock.

\begin{figure}
    \centering
    \includegraphics[width=0.95\textwidth]{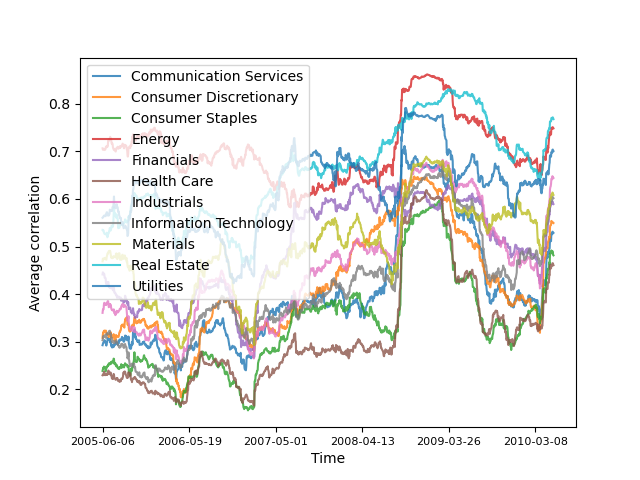}
    \caption{Average time-varying correlation for 11 sectors analyzed. Sectors considered include (a) Communication services, (b) Consumer Discretionary, (c) Consumer Staples, (d) Energy, (e) Financials, (f) Healthcare, (g) Industrials, (h) Information Technology, (i) Materials, (j) Real Estate, (k) Utilities. }
    \label{fig:Time_varying_correlation}
\end{figure}

Figure \ref{fig:Time_varying_correlation} displays the rolling average correlation exhibited by each equity sector. Some general patterns are clear. First, the ranking of various sectors' average correlation over time is remarkably consistent. That is, most sectors' average correlation trajectory appear to deviate from any other by some translation $\tau$, which is approximately constant over time. The second most notable insight is the obvious spike in average correlation levels during the GFC market crisis. This is consistent with a significant body of prior research, demonstrating that equity behaviours become more homogenous during market crises, with average correlations increasing. That being said, the degree of this increase varies significantly based on the sector - and this warrants further investigation.

The findings of our analysis are shown in Table \ref{tab:Average_sector_correlation}. There is a striking difference in the average correlation between difference sectors. During our period of analysis, the sector with the lowest average correlation is healthcare and the sector with the highest average correlation is energy. This is consistent with prior portfolio research, which has highlighted the typically low portfolio beta displayed by healthcare, and the high portfolio beta displayed by more volatile sectors such as energy. One interpretation of this analysis for those seeking to make stock selection decisions during periods of meaningful inflationary pressure, is the degree of homogeneity within candidate equity sectors. Sectors such as: Utilities, Real estate and Energy, all of which exhibit average correlation levels above $0.6$, display more homogenous behaviours and will provide less opportunities for relative alpha generation. By contrast, sectors such as Consumer Staples and Healthcare, both of which have average correlation below $0.35$, exhibit greater heterogeneity in equity behaviours. These sectors are likely to provide more opportunity for alpha generation during an environment of extreme inflationary pressure.


\section{Dynamic multi-asset optimization}
\label{Dynamic_multi_asset_optimisation}

We now explore a similar theme within the context of managing a portfolio of different asset classes. One notable development over the past several years has been the proliferation of digital currencies such as Bitcoin. Although investments in many coins such as Bitcoin and Ethereum provided very strong returns over the past several years, this was accompanied by significant volatility. But just how beneficial was cryptocurrency exposure to an investor focused on maximizing their risk-adjusted returns? We now try to address this pertinent and topical question. In particular, we study (in the context of managing a multi-asset financial portfolio), which asset classes would have provided the greatest utility to a core equity portfolio over the past $\sim$ 5 years? We take a collection of $a=1,...,A$ daily asset class returns where we let $A=7$, and study their behaviour over a date range from 01/01/2016 - 30/06/2021. We index these date $t_3=1,...,T_3$. The 7 securities studied include S\&P 500 equities, the Bloomberg LNG Index, London Metals Exchange, S\&P 500 energy index, CRB Commodity Index, Bloomberg REIT index and Bitcoin (as a representative of the cryptocurrency asset class). Let each asset classes' daily log returns be $R_a(t_3)$, where $t_3=1,...,T_3$ references a point in time $t_3$ during our window of analysis. We implement a time-varying portfolio optimization, where we use a rolling window of $250$ days, and study the resulting functions of time-varying optimal portfolio weights. We carefully select a $250$ day rolling window to simulate a portfolio manager who examines data over the past year to inform trading decisions. In the optimization problem below, we seek to learn portfolio weights $w_1,...,w_A$ such that we maximize the following

\begin{align}
\frac{\E (R_{p}(t_3)) - R_{f}}{\sigma^2_{p}(t_3)},  \forall t_3 \in \{251,...,T_3\}, \text{ where}\\
    \E (R_{p}) = \sum^{A}_{a=1} w_{a} R_{a} (t_3), \text{ and} \\
    \sigma^2_{p}(t_3) = \boldsymbol{w}(t_3)^{T} \Sigma(t_3) \boldsymbol{w}(t_3).
\end{align}

\begin{figure}
    \centering
    \includegraphics[width=0.95\textwidth]{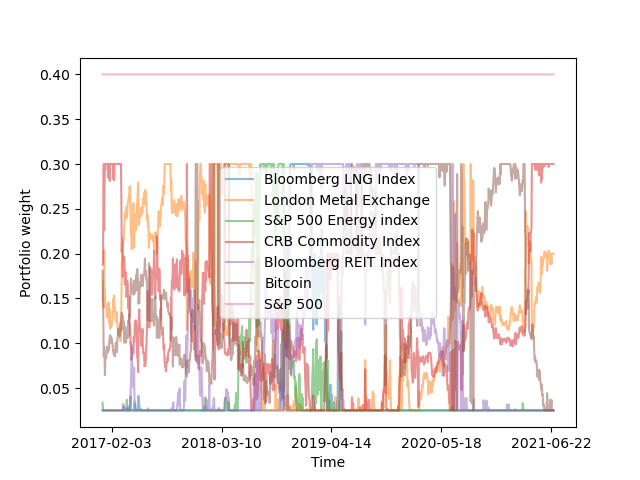}
    \caption{Time-varying optimal weight for each of our 7 candidate assets: (a) Bloomberg LNG Index, (b) London Metal Exchange, (c) S\&P 500 Energy Index, (d) CRB Commodity Index, (e) Bloomberg REIT Index, (f) Bitcoin, (g) S\&P 500.}
    \label{fig:Optimal_portfolio_weights}
\end{figure}

In the above expression, $\boldsymbol{w}(t_3)$ is a vector of portfolio weights at each point in time, $\Sigma_{ij}(t_3) = \text{cov}(R_{a,i}(t_3), R_{a,j}(t_3))$ is the time-varying covariance matrix between rolling log returns and $R_f$ is the current risk free rate which we set to 0.25\%. The constraints in this optimisation are as follows. We hold the S\&P 500 weight constant $a_1 = 0.4$, and restrict weights for the remaining portfolio assets $0.025 \leq w_{a} \leq 0.3, a = 2,...,A$ and assume that the portfolio is always fully invested in a long-only capacity $\sum^{A}_{a=1} w_{a} = 1$. At each point in time, the optimization generates a vector of optimal portfolio weights which we denote $w_a^{*} (t_3)$. For each asset class, we compute the mean and variance of the portfolio weights which we denote $\mu_a^{w^{*}}$ and $\sigma_a^{2, w^{*}}$. These results are shown in table \ref{fig:Optimal_portfolio_weights} which displays the mean and variance of the time-varying optimal portfolio weights among our collection of $A=7$ representative securities/indices. In conjunction with Figure \ref{fig:Optimal_portfolio_weights} which displays the optimal weight trajectories, there are several interesting findings.

\begin{table}
\centering
\begin{tabular}{ |p{3.8cm}||p{2.1cm}|p{2.1cm}|}
 \hline
 \multicolumn{3}{|c|}{Dynamic optimal portfolio weights} \\
 \hline
 Securities & $\mu_a^{w^{*}}$ & $\sigma_a^{2, w^{*}}$ \\
 \hline
 S\&P 500 equities & 0.4 & n/a \\
 LNG Index & 0.045 & 0.004 \\
 London Metals Exchange & 0.13 & 0.009 \\
 Energy index & 0.049 & 0.0045 \\
 CRB Commodity index & 0.116 & 0.0056 \\
 REIT index & 0.108 & 0.011 \\
 Bitcoin & 0.149 & 0.0088 \\
\hline
\end{tabular}
\caption{Mean and variance of time-varying optimal portfolio weight trajectories. Among our collection of more traditional hedging instruments, Bitcoin (cryptocurrency) was the most effective complement to a core equity portfolio in maximizing our portfolio's Sharpe Ratio.}
\label{tab:Optimal_portfolio_weight}
\end{table}

First, Bitcoin has the highest average optimal weight of 14.9\%. This suggests that for an investor with a core allocation to equities over the past 5.5 years, cryptocurrency was the most important asset in increasing the overall portfolio Sharpe Ratio. This was followed closely by the London Metals Exchange index with a total weight of 13\%. Although having an average optimal weight of 10.8\%, the REIT index exhibited the highest variance in its optimal weight over time. This suggests that REITs exposure may have to be monitored more closely than other asset classes, and may require more frequent rebalancing. Finally, although the LNG Index's average optimal allocation was only 4.5\%, it exhibited the lowest variance in its weight trajectory. This indicates that for a security exposure such as this, portfolio managers may require less trading and rebalancing. 


\section{Conclusion}
\label{Conclusion}

This paper uses various data, mathematical techniques and frameworks to provide a holistic view of inflation, and its impact on equity markets and (predominantly) equity market investors. We believe this is the first work to explore this topic from a macroeconomic, financial market dynamics and portfolio optimization perspective.

In Section \ref{Country_inflation_behaviours} we study the similarity of evolutionary CPI inflation patterns between a collection of 8 critically important economies. We demonstrate that Australia and Japan exhibited the most anomalous CPI trajectories, and this was primarily due to their hyper-inflationary onset preceding that of other major economies in the 1970s and 1980s. Our eigenvalue analysis demonstrated the strength of collective behaviours among country inflation patterns, reinforcing the systemic and largely ubiquitous nature of global inflation. We then introduce an inner-product based optimization framework to identify countries that are most central in global inflation. Japan is shown to be the least central (or most anomalous) country with respect to the timing of its inflationary patterns, with the majority of other countries forming one predominant cluster. Finally, we apply the Wasserstein distance between two conditional distributions of country equity index log returns to demonstrate that the USA and UK equity indices possess the most robustness during times of immense inflationary pressure.

In Section \ref{Equity_sector_dynamics} we partition S\&P 500 equities into 11 disjoint sectors, and study the evolutionary correlation behaviours during a period of extreme inflationary behaviours. Our analysis reveals significant differences between various equity sectors' correlation during times of inflationary crisis. Consumer staples and Healthcare exhibited the lowest average correlation of $0.34$ and $0.32$ respectively, while energy and real estate exhibited the highest average correlation among constituents with $0.7$ and $0.66$ respectively. Sectors with lower average correlation levels may produce more opportunities for alpha generation, relative to the remaining collection of equities, given the heterogeneity in dynamics. By contrast, sectors with higher average levels of correlation may provide less opportunities for portfolio alpha, with more homogeneity among the collection.

Finally in Section \ref{Dynamic_multi_asset_optimisation} we turn our attention to core equity investors over the past several years who are seeking favourable portfolio diversification. Among our collection of candidate asset classes one could use to diversify portfolio returns, cryptocurrency (Bitcoin) was identified as the most effective complement to a core exposure of equities. The second most effective asset class was the London Metals Exchange. In studying the resulting time-varying optimal weight trajectories, we computed the variance of each function as an indication as to how frequently asset classes may have to be rebalanced. Our analysis indicates that REITS exhibit the highest variance, and may have to be monitored most closely by portfolio managers.

There are several avenues for further work. First, one could explore similar questions using different types of mathematical and statistical analysis. One possible methodology for the study of country inflation trajectories could be the application of multiplex network analysis, for example. Second, one could apply the techniques introduced in this manuscript to answer more general macroeconomic questions, not focused on inflation specifically. Specific modelling techniques, such as our time-varying linear regression, could be replaced with moving average or autoregressive moving average models. Finally, our equity sector and portfolio optimization analysis could be applied on a more varied collection of underlying securities. Equity securities could include equities outside the USA, and the optimization section could include more asset classes, such as fixed income, for example.

\section{Appendix}
\label{Appendix}

\subsection{Optimization assets}
\label{Optimization_assets}

Below we include a list of the assets used in the optimization simulations.

\begin{table}
\centering
\begin{tabular}{ |p{5cm}||p{6.9cm}|}
 \hline
 \multicolumn{2}{|c|}{Asset definition} \\
 \hline
 Asset & Definition \\
 \hline
 S\&P 500 equity index (SPX) & Collection of 500 US equities. \\
 LNG Index (BCOMNG) & Bloomberg Natural Gas subindex. \\
 London Metals Exchange (LMEX) & Narrowly focused basket of industrial metals (aluminium, copper, nickel, zinc, tin, lead). \\
 Energy Index (S5ENRS) & S\&P 500 Energy index. \\
 CRB Commodity Index (CRB CMDT) & Index consisting of 19 commodity futures (eg. Oil, hogs, Cattle, Silver, Soybeans, etc.). \\
 REIT Index (BBREIT) & Collection of US REITs. \\
 Bitcoin & Bitcoin share price quoted in USD. \\
\hline
\end{tabular}
\caption{Assets used in Section 5 and their definition.}
\label{tab:Section_5_assets}
\end{table}

Below we test the robustness of our optimization finding under a variety of constant weights for the S\&P 500 equity index allocation. We set this weight equal to 0.2, 0.3, 0.4 and 0.5 in 4 different simulations. In all scenarios, cryptocurrency was the largest position in maximizing portfolio Sharpe Ratio. 

\begin{table}
\centering
\begin{tabular}{ |p{3.8cm}||p{2.1cm}|p{2.1cm}|p{2.1cm}|p{2.1cm}|}
 \hline
 \multicolumn{5}{|c|}{Dynamic optimal portfolio weights} \\
 \hline
 Securities & $\mu_a^{w^{*}}$ & $\mu_a^{w^{*}}$ & $\mu_a^{w^{*}}$ & $\mu_a^{w^{*}}$ \\
 \hline
 S\&P 500 equities & 0.2 & 0.3 & 0.4 & 0.5 \\
 LNG Index & 0.05 & 0.048 & 0.045 & 0.042 \\
 London Metals Exchange & 0.18 & 0.16 & 0.13 & 0.11 \\
 Energy index & 0.06 & 0.05 & 0.048 & 0.04 \\
 CRB Commodity index & 0.175 & 0.148 & 0.12 & 0.075 \\
 REIT index & 0.14 & 0.12 & 0.10 & 0.09 \\
 Bitcoin & 0.184 & 0.164 & 0.148 & 0.14 \\
\hline
\end{tabular}
\caption{Mean weight of portfolio assets under various assumptions for fixed S\&P 500 equity weight. Under all scenarios cryptocurrency (Bitcoin) was the most effective portfolio complement in maximizing the portfolio Sharpe Ratio.}
\label{tab:Optimal_portfolio_weight_various}
\end{table}

\begin{figure}
    \centering
    \begin{subfigure}[b]{0.48\textwidth}
        \includegraphics[width=\textwidth]{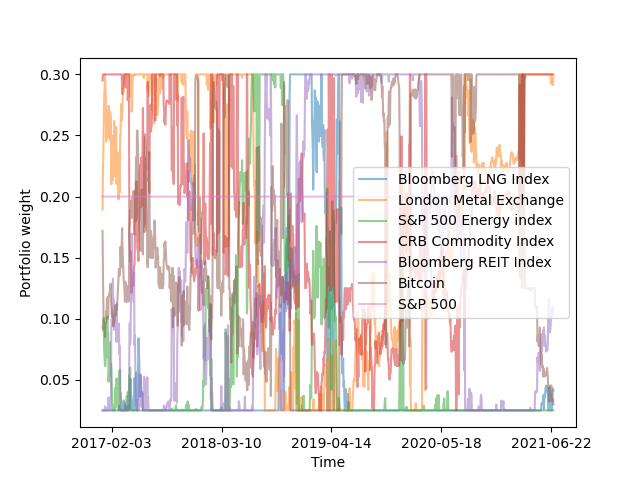}
        \caption{}
        \label{fig:Optimal_portfolio_0_2}
    \end{subfigure}
    \begin{subfigure}[b]{0.48\textwidth}
        \includegraphics[width=\textwidth]{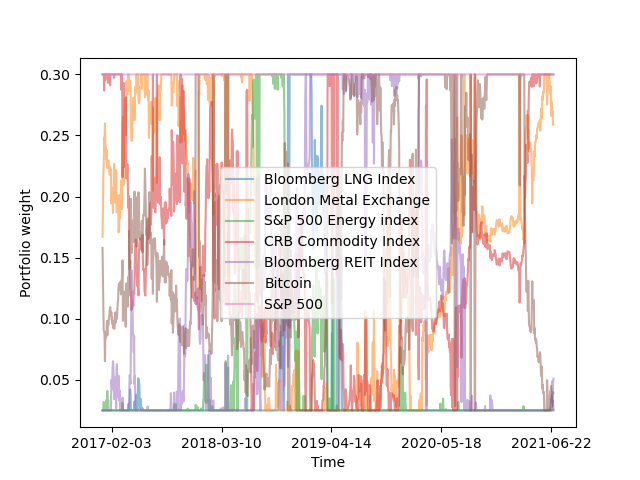}
        \caption{}
        \label{fig:Optimal_portfolio_0_3}
    \end{subfigure}
    \begin{subfigure}[b]{0.48\textwidth}
        \includegraphics[width=\textwidth]{Optimal_portfolio_weight_0_4.png}
        \caption{}
        \label{fig:Optimal_portfolio_0_4}
    \end{subfigure}
    \begin{subfigure}[b]{0.48\textwidth}
        \includegraphics[width=\textwidth]{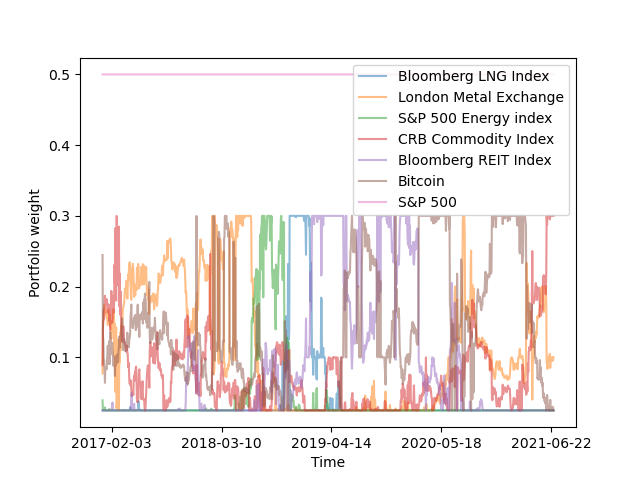}
        \caption{}
        \label{fig:Optimal_portfolio_0_5}
    \end{subfigure}
    \caption{Time-varying optimal weight for all 7 portfolio assets with varying levels of the constant S\&P 500 equity allocation. The S\&P 500 equity weight ranges in (a) 0.2, (b) 0.3, (c) 0.4 and (d) 0.5. }
    \label{fig:Optimal_portfolio_weight_appendix}
\end{figure}



\bibliographystyle{_elsarticle-num-names}
\bibliography{__References.bib}
\end{document}